# Towards organizational guidelines for the responsible use of AI

Richard Benjamins[1]

**Abstract.** In the past few years, several large companies have published ethical principles of Artificial Intelligence (AI). National governments, the European Commission, and inter-governmental organizations have come up with requirements to ensure the good use of AI. However, individual organizations that want to join this effort, are faced with many unsolved questions. This paper proposes guidelines for organizations committed to the responsible use of AI, but lack the required knowledge and experience. The guidelines consist of two parts: *i*) helping organizations to decide what principles to adopt, and *ii*) a methodology for implementing the principles in organizational processes. In case of future AI regulation, organizations following this approach will be well-prepared.

## 1    INTRODUCION

The popularization of AI has led to numerous applications such as content recommendation, chatbots, facial recognition, machine translation, fraud detection, medical diagnosis, etc. However, there are also risks associated to the massive uptake of AI such as unfair discrimination and opaque algorithmic decisions.

Those risks have motivated a range of organizations to come up with AI principles or ethics guidelines. The objective of this paper is to provide guidance to individual organizations in defining and implementing AI principles. Section 2 presents an overview of current principles. Section 3 proposes a three-step approach to zoom in on appropriate principles for an organization. Section 4 presents a methodology to implement the chosen principles into organizational structures. Finally, Section 5 provides conclusions.

## 2    PROLIFERATION OF AI PRINCIPLES

In the past three years, the amount of organizations publishing AI principles has grown significantly, including governments, private companies, civil societies, inter-governmental organizations and multi-stakeholder initiatives. There is general agreement on what principles are relevant for controlling AI. [1] gives an overview of more than thirty organizations with their respective principles classified into nine broad categories: human rights, human values, responsibility, human control, fairness & non-discrimination, transparency & explainability, safety & security, accountability, and privacy. Algorithm Watch maintains an open directory with AI principles of over 80 organizations [2], and [3] performed a global analysis of the AI principles of 84 organizations.

The European Commission has published its Ethics Guidelines for Trustworthy AI consisting of seven requirements [4]. Several governments have stated AI principles related to the future of work, liability of self-learning autonomous systems, malicious use, data monopolies & concentration of wealth. All in all, there is a large set of principles to choose from, yet there is little experience in what principles to choose and how to integrate them into organizational processes.

## 3    A THREE-STEP APPROACH TO FOCUS ON THE RELEVANT PRINCIPLES

Figure 1 provides an illustration of the many AI principles organizations can choose from. The following simple process can help to choose from the long list of principles.

1) **Distinguish** between principles relevant for **governments**, such as the future of work, lethal autonomous weapon systems, liability, concentration of power & wealth (right part of Figure 1), and principles that **individual organizations** (including private and public enterprises, public bodies, and civil societies) can act on, such as privacy, security, fairness and transparency (left part of Figure 1).

2) **Distinguish** between **intended and unintended** consequences. Many challenges of the use of AI are occurring as an unintended side effect of the technology (e.g. bias, lack of explainability, future of work, see top part of Figure 1). Intended consequences are explicit decisions and can be controlled, such using AI for good or for bad (bottom part of Figure 1). It is likely that with time, when organizations become more aware and capable of mitigating unintended consequences, those might become considered as intended if they continue to appear. Organizations better formulate their principles for the *unintended* consequences

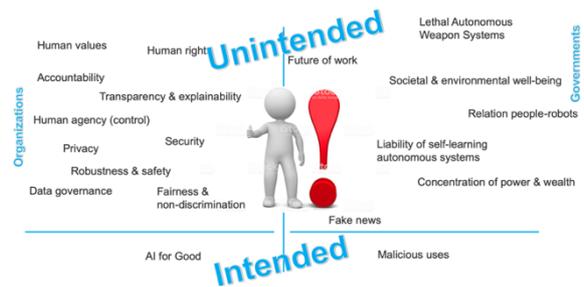

they can *act* upon (top left quadrant of Figure 1).

**Figure 1.** Classification of AI Principles along two dimensions: company-government (continuous) and unintended-intended.

[1] Telefonica, Spain, email: richard.benjamins@telefonica.com

3) Consider whether the AI Principles cover all aspects relevant for AI systems (e.g. safety, privacy, security, fairness, etc.) in an **end-to-end** manner, versus covering only **AI-specific** challenges (e.g. fairness, explainability, human agency). There is no hard line between those categories, but it is a continuum, as illustrated in Figure 2.

The decisions organizations take will be partly based on the **sector** they are in. For example, using AI in the aviation sector will put high value on safety, whereas the insurance sector will put high value on fairness and explainability.

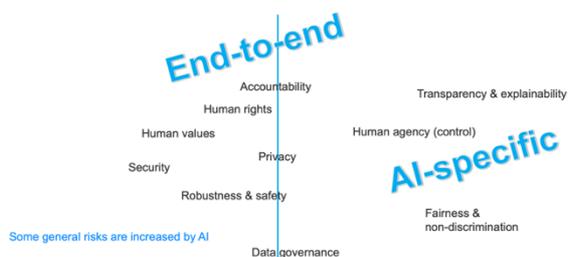

**Figure 2.** End-to-end principles versus AI-specific principles.

## 4 Responsible AI by Design

It is one thing to define the appropriate AI Principles, but it is another thing to make them part of "business as usual". In [5], we present such a methodology called "Responsible AI by Design". The methodology has five ingredients and is illustrated here with the case of Telefonica.

1) Telefonica's **AI Principles** state that the use of AI should be fair, transparent & explainable, human-centered, with privacy & security, which also applies to providers of AI solutions [6]. The Principles have been defined in a multi-departmental initiative based on existing literature and sectorial considerations.

2) It is important to provide **training** to employees explaining all relevant aspects. Figure 3 illustrates the modules of an online course developed by Telefonica. The content of the course is adapted to the technical savviness of employees.

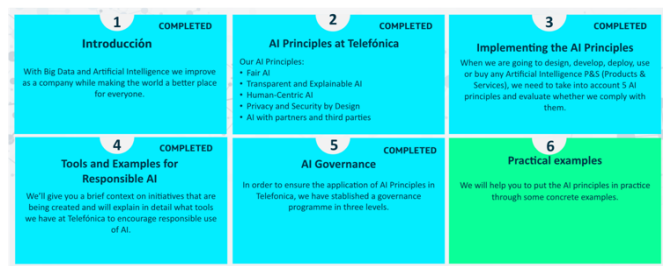

**Figure 3.** Modules of online AI Ethics training course for employees.

3) When designing, developing or buying AI systems, employees need to complete an online **questionnaire** with a set of questions and *recommendations*[2] corresponding to each principle. The questionnaire has been developed by AI experts in collaboration with Human Rights experts. All questionnaires are logged for governance reasons.

4) **Tools** are important for supporting automatic checking for bias in the data, mitigating potentially discriminatory algorithmic outcomes, finding proxy variables to sensitive variables, explainable AI for backbox algorithms, and data anonymization. Some of the tools are internally built while others are open source tools such as AI Fairness 360[3] and InterpretML[4].

5) A **governance model** defines responsibilities and the escalation process when the questionnaire reveals issues. We identified a new role called Responsible AI Champion for discussing potential issues. If not resolved, there are two levels of escalation: a multi-disciplinary team of experts, and our Responsible Business Office.

## 5 CONCLUSIONS

We have proposed an approach for the problem many organizations face when defining and implementing AI Principles. This problem has been recognized by the author in numerous discussions with organizations and is also evidenced by the pilot of the EC assessment list for trustworthy AI. Several AI experts of the HLEG[5] have confirmed the value of the presented approach for helping organizations move towards an ethical use of Artificial Intelligence.

---

[2] In contrast to the Assessment list for Trustworthy AI of the EC [4], that does not provide recommendations, but only asks questions.
[3] http://aif360.mybluemix.net/
[4] https://github.com/interpretml/interpret
[5] High-Level Expert Group on AI of the European Commission.